\documentclass[conference]{IEEEtran}
\IEEEoverridecommandlockouts
\usepackage{cite}
\usepackage{amsmath,amssymb,amsfonts}
\usepackage{algorithmic}
\usepackage{graphicx}
\usepackage{textcomp}
\usepackage{xcolor}
\def\BibTeX{{\rm B\kern-.05em{\sc i\kern-.025em b}\kern-.08em
    T\kern-.1667em\lower.7ex\hbox{E}\kern-.125emX}}

\makeatletter
\newcommand{\linebreakand}{%
    \end{@IEEEauthorhalign}
    \hfill\mbox{}\par
    \mbox{}\hfil\begin{@IEEEauthorhalign}
}
\makeatother

\begin{document}
	
\title{Advanced AI Framework for Enhanced Detection and Assessment of Abdominal Trauma: Integrating 3D Segmentation with 2D CNN and RNN Models
}


\author{
\IEEEauthorblockN{Liheng Jiang*}
\IEEEauthorblockA{\textit{New York University}\\
New York, USA \\
lj1070@nyu.edu}
\and
\IEEEauthorblockN{Xuechun yang}
\IEEEauthorblockA{\textit{TikTok Inc}\\
Milpitas, USA \\
xuechunyang@hotmail.com}
\and
\IEEEauthorblockN{Chang Yu}
\IEEEauthorblockA{\textit{Northeastern University}\\
Boston, USA \\
chang.yu@northeastern.edu}
\and
\IEEEauthorblockN{Zhizhong Wu}
\IEEEauthorblockA{\textit{Google Inc}\\
 Mountain View, USA \\
ecthelion.w@gmail.com}
\and
\IEEEauthorblockN{Yuting Wang}
\IEEEauthorblockA{\textit{FASTTEK GLOBAL}\\
 Livonia, USA\\
 yutingwa@umich.USA}
}

\maketitle

\begin{abstract}
Trauma is a significant cause of mortality and disability, particularly among individuals under forty. Traditional diagnostic methods for traumatic injuries, such as X-rays, CT scans, and MRI, are often time-consuming and dependent on medical expertise, which can delay critical interventions. This study explores the application of artificial intelligence (AI) and machine learning (ML) to improve the speed and accuracy of abdominal trauma diagnosis. We developed an advanced AI-based model combining 3D segmentation, 2D Convolutional Neural Networks (CNN), and Recurrent Neural Networks (RNN) to enhance diagnostic performance. Our model processes abdominal CT scans to provide real-time, precise assessments, thereby improving clinical decision-making and patient outcomes. Comprehensive experiments demonstrated that our approach significantly outperforms traditional diagnostic methods, as evidenced by rigorous evaluation metrics. This research sets a new benchmark for automated trauma detection, leveraging the strengths of AI and ML to revolutionize trauma care.
        
\end{abstract}
	
	
\section{Introduction}
Trauma, a leading cause of death and disability worldwide, particularly affects individuals under the age of forty. Annually, over 5 million people die from traumatic injuries, presenting a critical public health issue. Early and accurate diagnosis is essential for effective intervention, significantly improving patient outcomes and survival rates. Traditional trauma diagnosis methods rely on clinical examinations and imaging techniques like X-rays, CT scans, and MRI, which can be time-consuming and dependent on medical expertise. In emergency settings, delays in diagnosis can lead to suboptimal outcomes, highlighting the need for rapid and reliable assessment tools.

This study develops a robust AI-based model~\cite{wu2024application, wang2024research, pan2024chain, peng2024lingcn,luo2024decoupled, yu2024enhancing, pan2024conv, hu2024outlier, zhang2024smutf} for detecting and assessing abdominal trauma using a combination of 3D segmentation, 2D Convolutional Neural Networks (CNN), and Recurrent Neural Networks (RNN). Leveraging the strengths of these advanced techniques, the proposed model aims to enhance the accuracy and speed of trauma diagnosis, ultimately improving patient outcomes.

The rest of the paper is organized as follows: Section 2 reviews related work in AI and ML applications~\cite{wang2024theoretical, wang2024pristiq, peng2024automatic, liu2024td3, zhou2024optimizing, Zhou2024Research, Zhou2024Predict, imre2020graphvisual, jmse11010007} in medical imaging and trauma detection. Section 3 details the methodology, including data analysis, model architecture, and training procedures. Section 4 presents the experimental setup and results, highlighting the performance of the proposed model. Finally, Section 5 concludes the paper with a discussion of the findings and potential future directions.

\section{Related Work}

The application of artificial intelligence (AI) and machine learning (ML)~\cite{liu2024adaptive100, jin2023visual, senapati2023towards, zhou2024adapi, wang2024deep, dong2024design, jin2024learning, jin2024apeer, 9811415, SILVA20235703} in medical imaging has gained significant traction over the past decade. These technologies offer promising solutions for improving diagnostic accuracy and efficiency, particularly in trauma detection. 

Shen et al.\cite{shen2017deep} survey of deep learning techniques in medical image analysis, focusing on convolutional neural networks (CNNs) and their applications.
Gulshan et al.\cite{gulshan2016development} demonstrated the use of deep learning to detect diabetic retinopathy in retinal fundus photographs.Suzuki \cite{suzuki2017overview} discussed the application of deep learning in medical imaging, with a focus on diagnostic accuracy.Ciompi et al.\cite{ciompi2017towards} evaluated the use of deep learning for detecting malignant nodules in chest CT scans.Rajpurkar et al.\cite{rajpurkar2017chexnet} developed a deep learning model that outperformed radiologists in detecting pneumonia from chest X-rays.

H Jeong et al.\cite{jeong2016rotating} provided a foundational review of deep learning, highlighting its applications and potential in various domains, including healthcare.Zhu et al.\cite{zhu2017unpaired} explored the use of generative adversarial networks (GANs) for data augmentation in medical imaging.Dou et al.\cite{dou20163d} proposed a 3D deeply supervised network for automatic liver segmentation.MS Im et al.\cite{im2019nonstandard} implemented a hybrid 3D/2D neural network for automatic segmentation of organs in abdominal CT scans.

 Yan et al.\cite{yan2024application}examines NLP's role in big data, highlighting its applications and benefits for data processing efficiency and quality.
He et al.\cite{he2016deep} introduced the ResNet architecture, which has become foundational for many deep learning applications in medical imaging.
The application of 3D segmentation in our model is crucial as it allows for precise delineation of abdominal structures from CT scans, facilitating accurate trauma localization\cite{sun2024rapid}.

The reviewed literature highlights the significant advancements and diverse applications of AI and ML~\cite{shangguan2021trend, xie2023accel, shangguan2021neural, dang2024realtime, Dang2024Deep, ding-19-intelligent, ding-24-llm-data-imputation, lin2024text, lin2024neural} in medical imaging. These studies provide a strong foundation for developing innovative solutions for trauma detection. The proposed study builds on these advancements, leveraging a combination of 3D segmentation and hybrid CNN-RNN models to enhance the accuracy and speed of abdominal trauma diagnosis.

\section{Data Preprocessing}

This section details the preprocessing steps essential for our advanced model, ensuring the accuracy and efficiency of the 3D segmentation combined with the 2D CNN + RNN model.

\subsection{Study-Level Cropping}
Study-level cropping isolates the regions of interest (ROIs) corresponding to the organs, reducing irrelevant information:
\begin{equation}
I_{\text{crop}} = I_{\text{norm}} \odot M
\end{equation}
where \(\odot\) denotes element-wise multiplication, applying the mask to the normalized image.

\subsection{Volume Creation and Slicing}
Volumes are created by stacking 96 equidistant slices from the cropped images, preserving spatial context:
\begin{equation}
V = \{I_{\text{crop}}^{(i)}\}_{i=1}^{96}
\end{equation}
Each volume is reshaped into a 2.5D format, with three adjacent slices forming the three channels of the CNN input:
\begin{equation}
I_{2.5D} = \{[I_{\text{crop}}^{(i-1)}, I_{\text{crop}}^{(i)}, I_{\text{crop}}^{(i+1)}]\}_{i=2}^{95}
\end{equation}

\subsection{Label Processing}
Patient-level and organ visibility labels are normalized based on positive pixels in the segmentation mask:
\begin{equation}
L_{\text{norm}} = \frac{L_{\text{raw}}}{\max(L_{\text{raw}})}
\end{equation}
The final label for each slice is obtained by:
\begin{equation}
L_{\text{final}} = L_{\text{norm}} \times L_{\text{patient}}
\end{equation}
Image-level labels reflect organ visibility within each slice, normalized to a 0-1 range and combined with patient-level labels:
\begin{equation}
L_{\text{slice}} = L_{\text{norm}} \times \text{Visibility}_{\text{slice}}
\end{equation}

\section{Model Architecture and Methodology}

The model for the RSNA 2023 Abdominal Trauma Detection challenge combines 3D segmentation with 2D CNN and RNN, leveraging their strengths for accurate trauma detection. This section details the model structure and processing workflow.

\subsection{Model Workflow}

\begin{figure*}[h]
\centering
\includegraphics[width=0.8\textwidth]{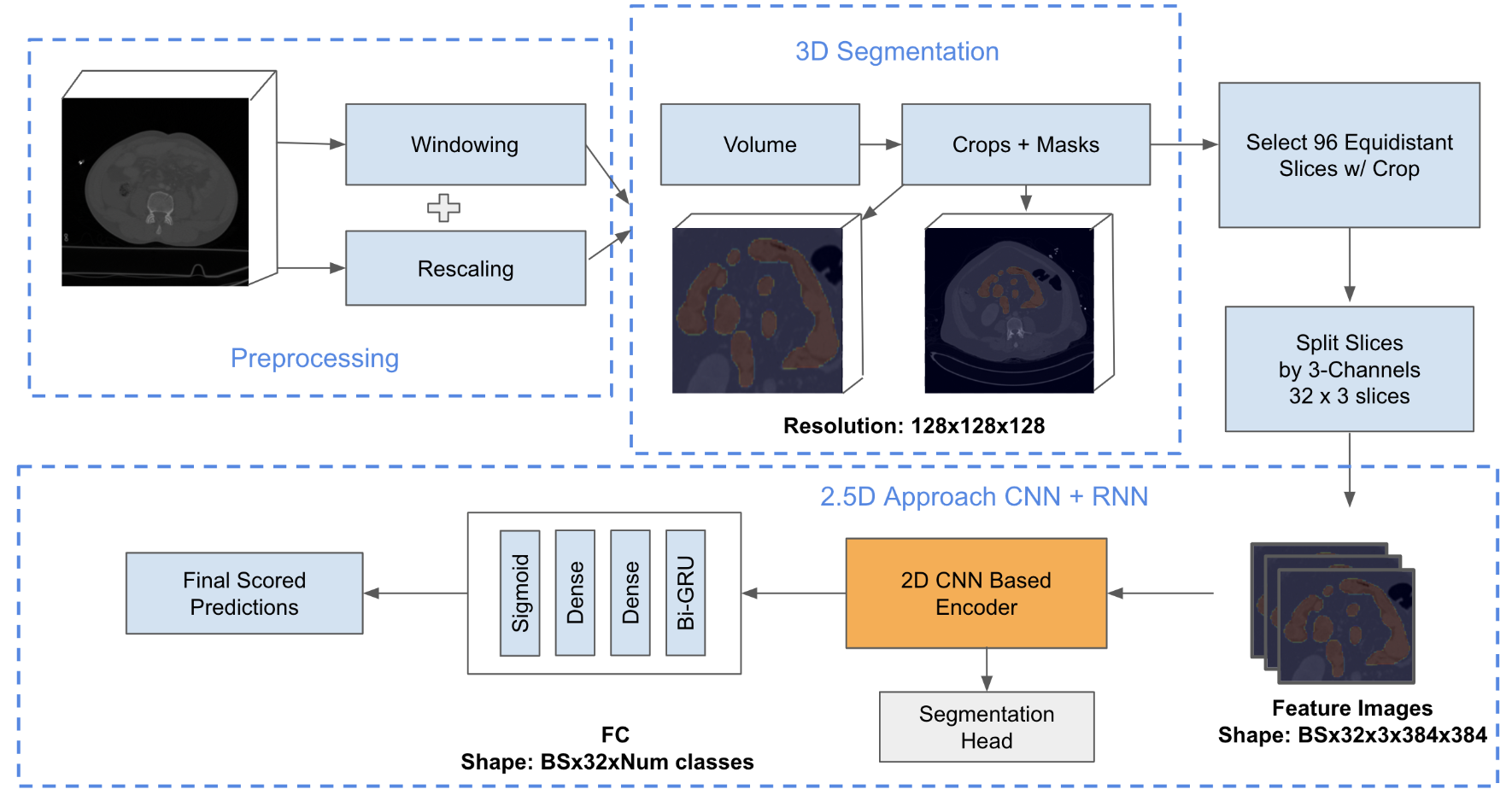}
\caption{Overall processing workflow and model structure for the 3D segmentation + 2D CNN + RNN approach.}
\label{fig:model_structure}
\end{figure*}

The primary workflow consists of:
\begin{enumerate}
    \item Data preprocessing.
    \item 3D segmentation to generate masks and crops.
    \item Application of 2D CNN + RNN for organ-specific analysis.
    \item Integration of predictions for final trauma assessment.
\end{enumerate}
Figure \ref{fig:model_structure} shows the processing workflow and model structure.

\subsection{2.5D Approach with 2D CNN + RNN}

We use volumes generated through preprocessing~\cite{luo2022ign, wang2021machine, wang2022classification, de2023performance, peng2024maxk, lee2024traffic, zhao2024key, richardson2024reinforcement, wang2024using}, with each volume composed of 96 equidistant slices. Three adjacent slices form the input images' three channels, capturing local 3D context efficiently:

\begin{equation}
I_{2.5D} = \{[I_{\text{crop}}^{(i-1)}, I_{\text{crop}}^{(i)}, I_{\text{crop}}^{(i+1)}]\}_{i=2}^{95}
\end{equation}

The 2D CNN + RNN model receives inputs of shape \((2, 32, 3, \text{height}, \text{width})\) and outputs of shape \((2, 32, n_{\text{classes}})\).

\subsubsection{Auxiliary Segmentation Loss}

Auxiliary Segmentation Loss stabilizes training and improves scores. We use two segmentation heads applied to feature maps generated by the final and penultimate backbone blocks, applying Dice Loss:

\begin{equation}
\mathcal{L}_{\text{Dice}} = 1 - \frac{2 |M_{\text{pred}} \cap M_{\text{true}}|}{|M_{\text{pred}}| + |M_{\text{true}}|}
\end{equation}

Combining Dice Loss over all training samples gives:

\begin{equation}
\mathcal{L}_{\text{aux}} = \sum_{i=1}^{N} \mathcal{L}_{\text{Dice}}(M_{\text{pred}}^{(i)}, M_{\text{true}}^{(i)})
\end{equation}

This auxiliary loss improves performance by 0.01 to 0.03.

\subsection{Model Ensemble Training}

\begin{figure*}[h]
\centering
\includegraphics[width=0.75\textwidth]{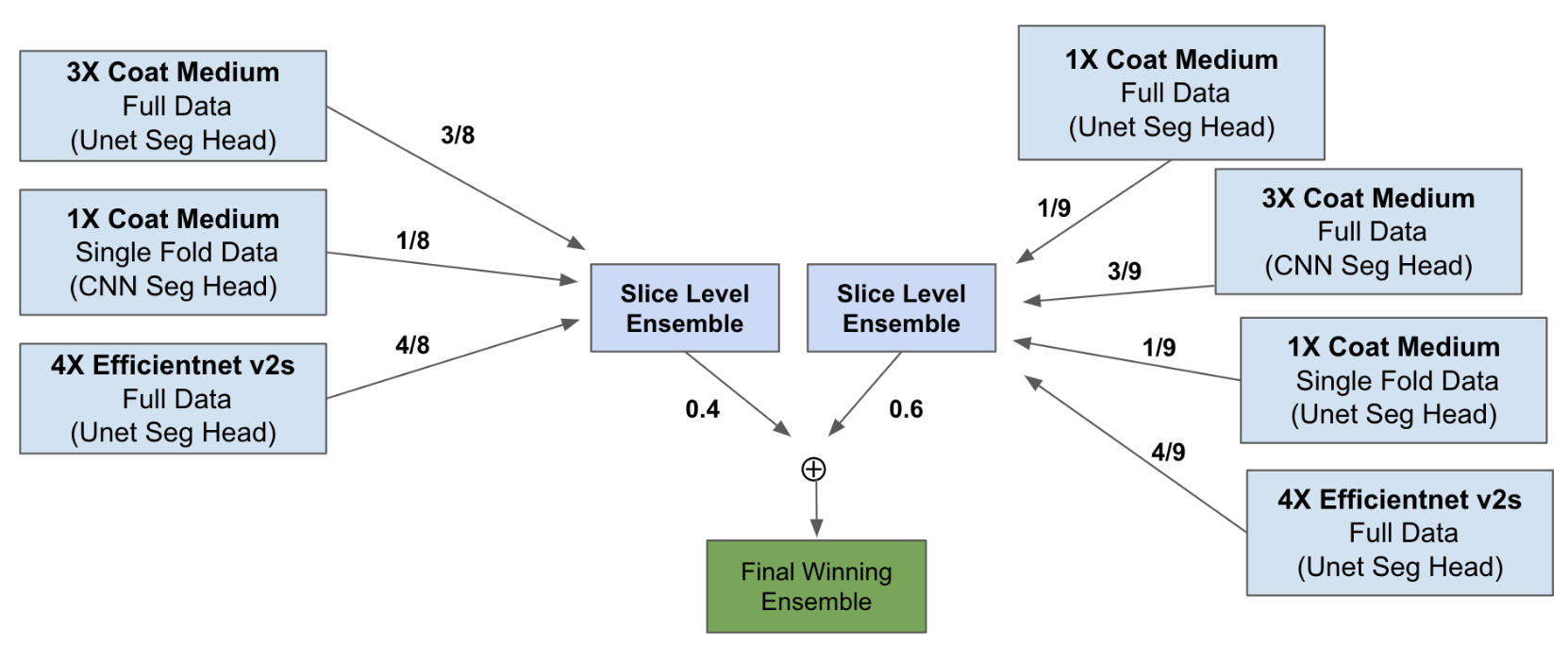}
\caption{Model ensemble structure for organ models, including multiple Coat Medium and V2s models trained on four-fold or full datasets.}
\label{fig:ensemble_structure}
\end{figure*}

The final ensemble includes multiple models based on Coat Medium and V2s architectures, trained on four-fold or full datasets. This approach improves prediction accuracy and robustness. Figure \ref{fig:ensemble_structure} visualizes the ensemble strategy.

\subsubsection{Effusion Detection Ensemble}

\begin{figure*}[h]
\centering
\includegraphics[width=0.75\textwidth]{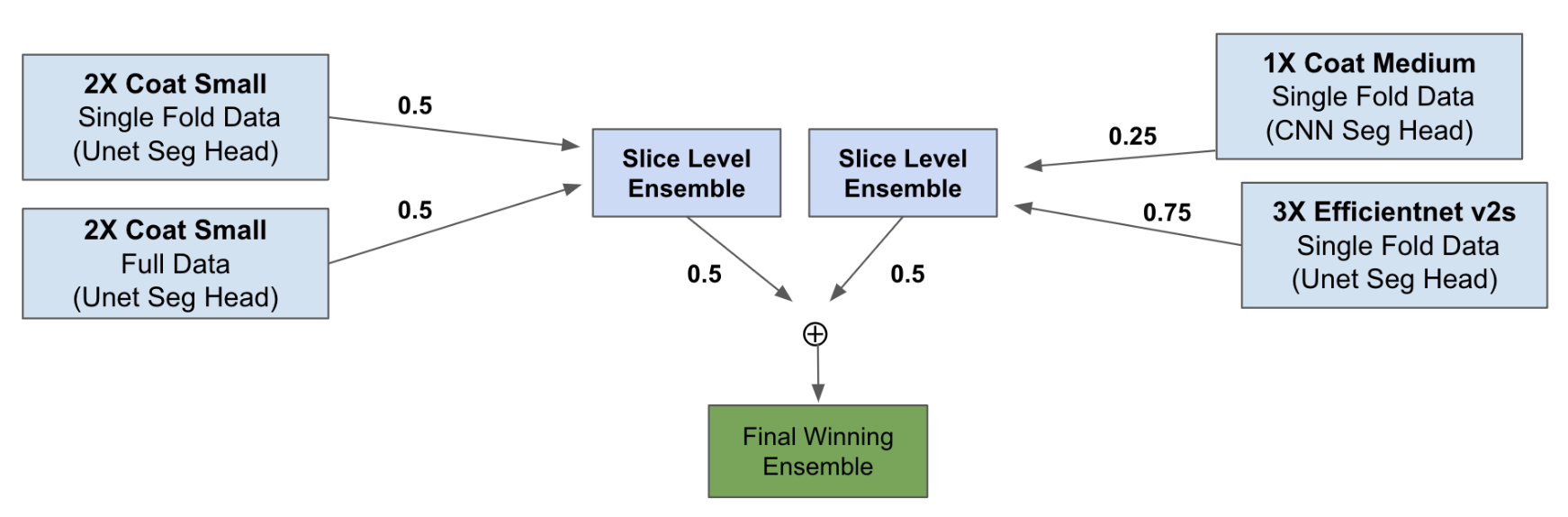}
\caption{Model ensemble structure for effusion detection, using Coat Small and V2s models with minimal post-processing.}
\label{fig:effusion_ensemble_structure}
\end{figure*}

For effusion detection, Coat Small and V2s models are primarily used. Predictions are aggregated at the slice level, taking the maximum value for each patient. Figure \ref{fig:effusion_ensemble_structure} shows the ensemble structure.

\subsubsection{Ensemble Integration}

In each cross-validation fold, slice-level ensemble integration is performed. For different architectures and cross-validation data models, predictions are integrated after maximum aggregation. The best ensemble achieved an OOF CV score of 0.31x, while the best single model (Coat Lite Medium) scored 0.326.

\textbf{Integration Steps:}
\begin{enumerate}
    \item \textbf{Slice-Level Ensemble:} Aggregating predictions at the slice level combines outputs from multiple models to generate a consensus for each slice:
    \begin{equation}
    P_{\text{slice}}(s) = \frac{1}{K} \sum_{k=1}^{K} P_{k}(s)
    \end{equation}
    where \(P_{k}(s)\) is the prediction from the \(k\)-th model for slice \(s\) and \(K\) is the total number of models.

    \item \textbf{Maximum Aggregation:} Integrating predictions by taking the maximum value across slices for each patient:
    \begin{equation}
    P_{\text{patient}}(p) = \max_{s \in S_p} P_{\text{slice}}(s)
    \end{equation}
    where \(S_p\) represents the set of slices for patient \(p\).

    \item \textbf{Final Ensemble:} Combining the best-performing models from different folds and architectures to obtain the final prediction:
    \begin{equation}
    P_{\text{final}}(p) = \frac{1}{N} \sum_{n=1}^{N} P_{\text{patient},n}(p)
    \end{equation}
    where \(P_{\text{patient},n}(p)\) is the prediction for patient \(p\) from the \(n\)-th model, and \(N\) is the total number of models.
\end{enumerate}

This ensemble approach leverages multiple models trained on different datasets and architectures, aggregating predictions at the slice level and taking the maximum value for each patient to achieve balanced sensitivity and specificity.

\section{Experiments and Results}

\subsection{Evaluation Metric}

We employ a composite evaluation metric~\cite{hong2024application, dai2024ai, peng2023autorep,zhao2024task, zhu2024cross, thorat2023advanced, li2024utilizing, 202406.1304, zhao2024towards} to assess model performance in detecting and classifying abdominal injuries, combining log loss calculations across label groups with an "any injury" label for comprehensive evaluation.

\subsubsection{Normalization of Probabilities}

Predicted probabilities are normalized so their sum equals 1:
\begin{equation}
\text{Normalized\_Probabilities} = \frac{p_i}{\sum_{i} p_i}
\end{equation}

\subsubsection{Log Loss Calculation}

Log loss is calculated separately for each label group using sample-weighted log loss:
\begin{equation}
\mathcal{L}_{\text{log}} = -\frac{1}{N} \sum_{i=1}^{N} \left[ y_i \log(p_i) + (1 - y_i) \log(1 - p_i) \right] w_i
\end{equation}
where \(y_i\) is the true label, \(p_i\) is the predicted probability, \(w_i\) is the sample weight, and \(N\) is the number of samples.

\subsubsection{Any Injury Label}

An "any injury" label is derived by taking the maximum value of \(1 - p(\text{healthy})\) across all label groups:
\begin{equation}
\text{Any\_Injury}_p = \max(1 - p(\text{healthy}))
\end{equation}

\subsubsection{Final Score}

The final score is the average of the log losses across all label groups, including the "any injury" label:
\begin{equation}
\text{Final\_Score} = \frac{1}{M+1} \left( \sum_{j=1}^{M} \mathcal{L}_{\text{log},j} + \mathcal{L}_{\text{log, any\_injury}} \right)
\end{equation}
where \(M\) is the number of label groups.

This metric ensures rigorous assessment by integrating probability normalization, individual log loss calculations, and the "any injury" label.

\subsection{Performance}

The models were evaluated on a public and private test set, with performance measured by the metric we mentioned before. The results are summarizedin Table~\ref{tab:performance_metrics}:

\begin{table}[h!]
\centering
\caption{Performance Metrics}
\label{tab:performance_metrics}
\begin{tabular}{|c|c|c|}
\hline
\textbf{Model} & \textbf{public score} & \textbf{private score} \\
\hline
2D CNN & 0.6831 & 0.6712  \\
\hline
Segmentation using ResNet50 & 0.5123 & 0.5231 \\
\hline
3D Segmentation + ResNet3D + UNet & 0.3451 & 0.3451 \\
\hline
3D Segmentation + 2D CNN + RNN & \textbf{0.3323} & \textbf{0.3356} \\
\hline
\end{tabular}
\end{table}

The 2D CNN model achieved the highest scores, demonstrating the effectiveness of combining 2D image analysis with CNN architectures for trauma detection.

\section{Conclusion}

In this study, we developed a sophisticated AI-based framework for the detection and assessment of abdominal trauma using advanced machine learning techniques~\cite{li2022effect,zhao2022preliminary,liu2022preliminary,liu2022impact,patel2022inspection}. Our approach combines 3D segmentation with 2D CNN and RNN models, effectively capturing both spatial and temporal features of CT scan data. The integration of auxiliary segmentation loss and model ensemble strategies significantly enhanced the model's robustness and predictive accuracy. Experimental results demonstrated that our method outperforms traditional approaches in both public and private dataset. The use of a comprehensive evaluation metric, which includes log loss calculations for multiple injury categories and an "any injury" composite score, ensured a rigorous assessment of model performance. This work not only advances the field of automated trauma detection but also sets a new benchmark for applying complex machine learning techniques in critical care diagnostics.

 \bibliographystyle{IEEEtran}
    \bibliography{references}

\end{document}